\documentclass[journal=jacsat,manuscript=article]{achemso}

\usepackage{multicol}
\usepackage{graphicx}
\usepackage{chemformula} % Formula subscripts using \ch{}
\usepackage[utf8]{inputenc}
\DeclareUnicodeCharacter{0308}{\"{}}
\usepackage[T1]{fontenc} % Use modern font encodings

\author{Jane Elisa Guimarães}
\affiliation[UFMG]
{Departamento de Física, Universidade Federal de Minas Gerais, Belo Horizonte, MG, 31270-901, Brazil}

\author{Rafael Nadas}
\affiliation[Berlin]
{Institut f\"{u}r Physik, Humboldt-Universit\"{a}t zu Berlin, Newtonstraße 15, 12489, Berlin, Germany}

\author{Rayan Alves}
\affiliation[UFMG]
{Departamento de Física, Universidade Federal de Minas Gerais, Belo Horizonte, MG, 31270-901, Brazil}

\author{Wenjin Zhang}
\affiliation[Tokyo]
{Department of Physics, Tokyo Metropolitan University, Tokyo, Japan}
\alsoaffiliation[NIMSnanoarchi]
{Research Center for Materials Nanoarchitectonics, National Institute for Materials Science, Tsukuba, Japan}

\author{Takahiko Endo}
\affiliation[Tokyo]
{Department of Physics, Tokyo Metropolitan University, Tokyo, Japan}
\alsoaffiliation[NIMSnanoarchi]
{Research Center for Materials Nanoarchitectonics, National Institute for Materials Science, Tsukuba, Japan}

\author{Kenji Watanabe}
\affiliation[NIMS]
{Research Center for Electronic and Optical Materials, National Institute for Materials Science, Tsukuba, Japan}

\author{Takashi Taniguchi}
\affiliation[NIMSnanoarchi]
{Research Center for Materials Nanoarchitectonics, National Institute for Materials Science, Tsukuba, Japan}

\author{Riichiro Saito}
\affiliation[Tokyo]
{Department of Physics, Tokyo Metropolitan University, Tokyo, Japan}
\alsoaffiliation[Tohoku]
{Department of Physics, Tohoku University, Sendai, Japan}

\author{Yasumitsu Miyata}
\affiliation[Tokyo]
{Department of Physics, Tokyo Metropolitan University, Tokyo, Japan}
\alsoaffiliation[NIMSnanoarchi]
{Research Center for Materials Nanoarchitectonics, National Institute for Materials Science, Tsukuba, Japan}

\author{Bernardo R. A. Neves}
\affiliation[UFMG]
{Departamento de Física, Universidade Federal de Minas Gerais, Belo Horizonte, MG, 31270-901, Brazil}

\author{Ado Jorio}
\affiliation[UFMG]
{Departamento de Física, Universidade Federal de Minas Gerais, Belo Horizonte, MG, 31270-901, Brazil}
\email{adojorio@fisica.ufmg.br}

\title[]
  {Nano-Raman Spectroscopy Analysis of Nanoprotuberances in MoSe$_2$}

\abbreviations{TERS,AFM,TMDC,SPM,PTTP,CVD,PCA}

\begin{document}


\begin{abstract}
Contaminations in the formation of two-dimensional heterostructures can hinder or generate desired properties. Recent advancements have highlighted the potential of tip-enhanced Raman spectroscopy (TERS) for studying materials in the 2D semiconductor class. In this work, we investigate the influence of 50-200nm sized nanoprotuberances within a monolayer of MoSe$_2$ deposited on hBN using nano-Raman spectroscopy, establishing correlations between the presence of localized contaminations and the observed hyperspectral variations. A figure of merit is established for the identification of surface impurities, based on MoSe$_2$ peaks ratio. Notably, new spectral peaks were identified, which are associated with the presence of nanoprotuberances and may indicate contamination and oxidation.
\end{abstract}


% For two cols:
%\begin{multicols}{2}
%%%%%%%%%%%%%%%%%%%%%%%%%%%%%%%%%%%%%%%%%%%%%%%

\section{Introduction}

Transition-metal dichalcogenides (TMDCs), such as MoSe$_2$, are increasingly studied due to their semiconducting properties and potential in optoelectronic applications \cite{wang2012electronics, ji2016strain}. These materials are particularly valuable for enhancing device performance, making it essential to investigate their vibrational behaviors, excitonic interactions, and strain effects.

Van der Waals heterostructures, created by stacking thin atomic layers of different materials, enable the combination of materials with complementary properties. Although these structures foster strong adhesion between adjacent two-dimensional materials, contamination can be present on the surfaces of each layer prior to assembly. This contamination, which may include adsorbed water molecules, oxidation, or hydrocarbons, is usually confined to nanoscale protuberances \cite{haigh2012cross,khestanova2016universal}. When these protuberances are formed by trapped gases or droplets, they are commonly called nanobubbles \cite{attard2002nanobubbles,lohse2015surface}. 

Tip-enhanced Raman spectroscopy (TERS) is a material characterization technique that provides nanometric resolution for nano-Raman spectroscopy \cite{Stockle_2000, Hayazawa_2000, Anderson2000}. The maximum spatial resolution achievable with conventional optical microscopy is approximately half of the excitation wavelength. TERS was developed to overcome this limitation \cite{kumar2015,jorio20242d,hoppener2024tip}, by combining scanning probe microscopy (SPM) with Raman spectroscopy. By employing a metallic tip to locally amplify the electromagnetic field, the technique provides enhancement of the Raman signal and improvement of spatial resolution, enabling detailed characterization of materials at the nanometer scale \cite{Vasconcelos_2018}.

\begin{figure}[t]
\centering
\includegraphics[width=\textwidth]{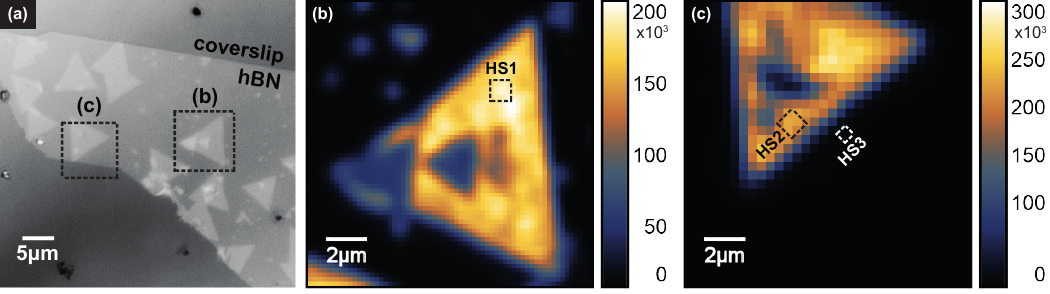}
\caption{(a) Optical microscopy image of the MoSe$_2$/hBN sample, showing regions of monolayer and bilayer over an hBN flake. The highlighted triangles correspond to the areas where different nano-Raman hyperspectra were measured, indicated in (b,c). (b) Avalanche photodiode (APD) micro-Raman map of the MoSe$_2$ heterostructure where the HS1 nano-Raman hyperspectrum was acquired. (c) APD micro-Raman map of the MoSe$_2$ heterostructure where the HS2 and HS3 nano-Raman hyperspectra were acquired. The colored scale bars in (b,c) indicate the MoSe$_2$ confocal photoluminescence (PL) intensity (photo counts in arbitrary units).}
\label{fig:APD}
\end{figure}

In this study, nano-Raman hyperspectral measurements have been used to investigate nanoscale protuberances on TMDCs heterostructures, more specifically MoSe$_2$ monolayer on an hBN substrate (Figure \ref{fig:APD}). These protuberances can induce localized variations in strain, dielectric environment, and charge distribution, potentially altering the vibrational modes and optical responses of surfaces \cite{lloyd2016band, darlington2020facile}. Here, we define a figure of merit for surface contamination based on the intensity ratios of MoSe$_2$ peaks. Additionally, new peaks, unrelated to the intrinsic properties of the TMDCs, are here shown to emerge due to trapped contaminants within nanoprotuberances or molecules responsible for their formation.

\section{Results}

\subsection{Atomic Force Microscopy}

\begin{figure}[ht]
\centering
\includegraphics[width=0.55\textwidth]{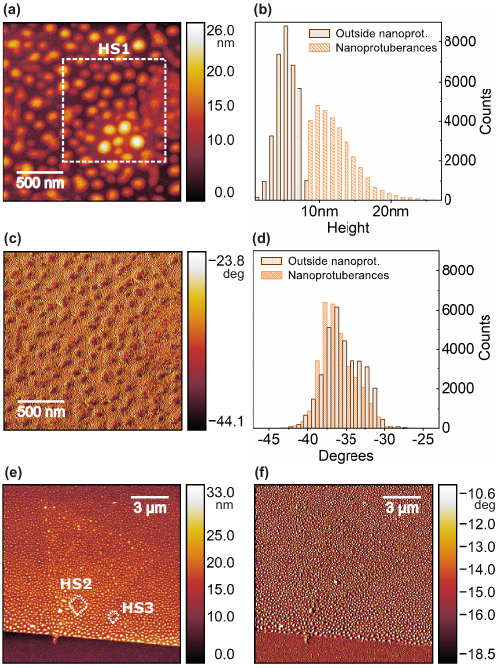}
\caption{(a) Topographic map of the MoSe$_2$ sample surface, revealing nanoscale protuberances measured by atomic force microscopy (AFM). The HS1 area is highlighted. (b) Histogram of height distribution, counted by pixels distinguished by regions inside and outside the nanoprotuberances. (c) Corresponding phase map. (d) Histogram of phase contrast distribution. (e) Topographic map of the hBN edge with the MoSe$_2$ flake, highlighting the HS2 and HS3 regions. (f) Corresponding phase map.}
\label{fig:AFM}
\end{figure}

Nanosized protuberances were observed in the MoSe$_2$ sample through atomic force microscopy (AFM) maps, with diameters ranging from 50 to 200nm and heights ranging from 8 to 25nm, as shown in Figures \ref{fig:AFM}(a,b). This region corresponds to hyperspectral data HS1. They are present in both the MoSe$_2$ region and the hBN-only region, as shown in Figure \ref{fig:AFM}(e), which displays the coverslip/hBN and hBN/MoSe$_2$ interfaces.

In addition to topography, phase contrast maps are acquired during tapping-mode AFM. Phase shifts indicate the difference between excitation and response of the cantilever, and that contrast enables the differentiation of surface features at the nanoscale, indicating energy dissipation associated with tip-sample interactions \cite{garcia1998phase, garcia1999attractive,phani2021deconvolution}. The maps in Figures \ref{fig:AFM}(c,f)) reveal phase contrasts when transitioning from protuberance regions to surrounding areas. Additionally, a clearer contrast is observed between the coverslip and hBN, whereas the contrast at the hBN/MoSe$_2$ interface is subtle. The histogram in Figure \ref{fig:AFM}(d) indicates two distinct peaks in the phase distribution, distinguishing nanoprotuberance regions from flat areas. It is worth noting that the phase values are qualitative for the equipment used. By correlating the topography and phase contrast maps, the observed heterogeneous pattern can be attributed to the presence of contaminant between the material layers or on top of the MoSe$_2$ flake.

\subsection{TERS figure of merit for MoSe$_2$ contamination}

Nano-Raman hyperspectral analyses were acquired from three distinct regions of the sample, as indicated in the APD images shown in Figure \ref{fig:APD}. Two hyperspectral maps (HS1 and HS2) were collected from an area where MoSe$_2$ is over hBN (MoSe$_2$/hBN), while the third one (HS3) was obtained from a region consisting solely of hBN. 

\begin{figure}[h]
\centering
\includegraphics[width=\textwidth]{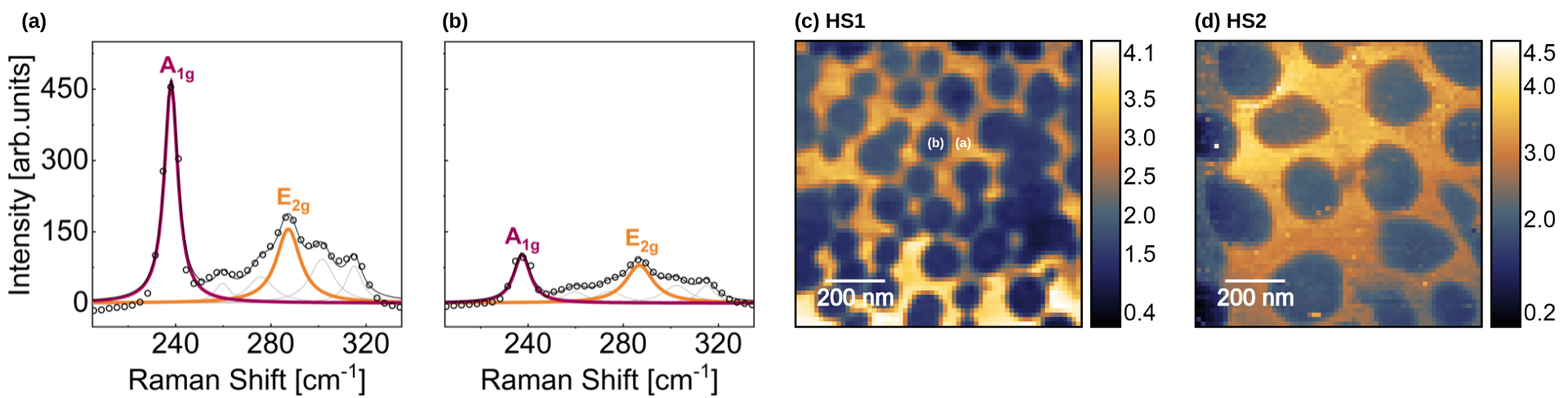}
\caption{Comparison of MoSe$_2$ characteristic TERS spectra (a) outside and (b) inside nanoprotuberances, corresponding respectively to regions (a) and (b) marked in (c). Intensity ratio maps of the MoSe$_2$ A$_{1g}$ band to the E$_{2g}$ band for (c) HS1 and (d) HS2 hyperspectra. The color-coded scale bars represent intensity ratios, ranging from lower values (blue regions) to higher values (orange regions).}
\label{fig:A1gdivE2g}
\end{figure}

On top of the nanoprotuberances, the intensity ratio of the A$_{1g}$ band of MoSe$_2$ — 240 cm$^{-1}$ — to the E$_{2g}$ band — 287 cm$^{-1}$ — distinguishes itself from flat regions across both maps acquired in MoSe$_2$ regions (HS1 and HS2), as illustrated in Figure \ref{fig:A1gdivE2g}. This provides a figure of merit for spectrally differentiating regions inside and outside the protuberances based on the TERS hyperspectral data. 

In a far-field map acquired in the same region as HS1, shown in Figure S1 (Supporting Information), the mean intensity ratio $I_{A_{1g}}/I_{E_{2g}}$ for the 4096 acquired spectra is (1.47 ± 0.08). A similar average analysis was performed for the two near-field maps, distinguishing between regions inside and outside the protuberances, an approach not feasible in the far-field due to the inability to spatially resolve regions of protuberance without the engaged TERS tip. Data were masked to separately analyze protuberances and their surroundings, given that in the two maps, areas of protuberance correspond to approximately 50\% of the total mapped region. The mean intensity ratios are shown in Table \ref{table2}.

\begin{table}[ht]
    \centering

        \begin{tabular}{lccc}
            \hline
            \textbf{Map} & \textbf{Nanoprotuberances} & \textbf{Surroundings} \\
            \hline
            \textbf{HS1}  & (1.5 ± 0.4) & (2.8 ± 0.8)  \\
            \textbf{HS2}  & (1.7 ± 0.3) & (2.5 ± 0.5)  \\
            \hline
        \end{tabular}

    \caption{Mean intensity ratios $(A_{1g}/E_{2g})$ for different areas in HS1 and HS2.}
    \label{table2}
\end{table}

The results reinforce that the intensity ratio is lower in the near-field regime over nanoprotuberances, where the enhancement of the A$_{1g}$ mode is less pronounced compared to the surrounding flat areas. This lower enhancement of the A$_{1g}$ mode (when compared to the E$_{2g}$ mode) can be attributed to the loss of light coherence in the near-field, as discussed in references ~\citenum{Cancado2014,jorio20242d,nadas2025tip}. Another contributing factor might be the slight increase in tip–sample separation caused by the small surface features on top of the MoSe$_2$ flake. This increase in tip-sample distance is known to affect the optical signal in near-field measurements \cite{novotny2012nanooptics}.

For protuberant areas, the mean intensity ratios found were (1.5 ± 0.4) and (1.7 ± 0.3) for the hyperspectral maps HS1 and HS2, respectively. In contrast, for regions outside the protuberances, the corresponding values were (2.8 ± 0.8) and (2.5 ± 0.5). These findings suggest that the strong enhancement of the A$_{1g}$ mode observed in the near-field regime does not propagate to the far-field. 

\subsection{Chemical analysis of contaminants}

\begin{figure}[H]
\centering
\includegraphics[width=0.6\textwidth]{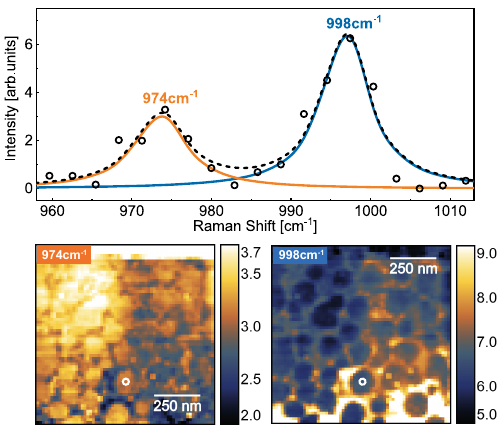}
\caption{Curve fitting for characteristic peaks in nanoprotuberance regions and corresponding TERS intensity maps for HS1 area. The orange Lorentzian corresponds to the peak at 974cm$^{-1}$ and the blue to 998cm$^{-1}$. Experimental data are represented by open black circles, while the dashed line corresponds to the final fitted curve. The white circle indicates the pixel in the map from which the spectrum data was extracted.}
\label{fig:MoSe2_1}
\end{figure}

In addition to MoSe$_2$ characteristic modes, distinct Raman peaks were observed in the regions of protuberances, indicating chemical species unique to these areas. We begin our analysis with hyperspectrum HS1, with intensity maps for two of the distinct peaks shown in Figure \ref{fig:MoSe2_1}. In HS1, a characteristic peak related to the presence of protuberances is observed at 974 cm$^{-1}$. When the amplitude of the Lorentzian fit corresponding to this peak is mapped, the nanoprotuberances become prominent, as the amplitude is higher in these regions. In contrast, the Lorentzian associated with the 998 cm$^{-1}$ peak exhibits an inverse spatial distribution, with higher intensity in the regions surrounding the protuberances. The 998 cm$^{-1}$ peak is likely associated with MoO$_3$ \cite{krishna2016investigation, kothaplamoottil2019greener}, suggesting possible oxidation of MoSe$_2$ across the sample. The presence of the lower-frequency peak within the protuberances, close to the known vibrational frequency of MoO$_3$, may indicate local charge doping effects in these regions. Alternatively, this may reflect two independent peaks, which nonetheless still suggest the presence of oxidation.

\begin{figure}[H]
\centering
\includegraphics[width=0.6\textwidth]{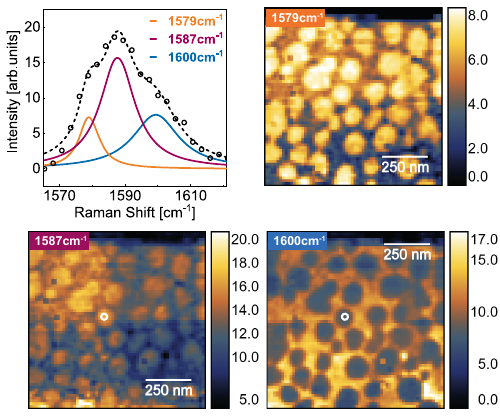}
\caption{Curve fitting for characteristic peaks in nanoprotuberance regions and corresponding TERS intensity maps for HS1 area. The orange Lorentzian corresponds to the peak at 1579cm$^{-1}$, the magenta to 1587cm$^{-1}$ and the blue to 1600cm$^{-1}$. Experimental data are represented by open black circles, while the dashed line corresponds to the final fitted curve. The white circle indicates the pixel in the map from which the spectrum data was extracted.}
\label{fig:MoSe2_2}
\end{figure}

In Figure \ref{fig:MoSe2_2}, three Lorentzian fits are shown along with their corresponding amplitude distributions across the HS1 map. The most intense peak, located at 1585 cm$^{-1}$, can be attributed to the C=C stretching mode, suggesting the presence of carbon-based contamination \cite{jorio2011raman}. The spatial distribution of the higher-frequency Lorentzian highlights regions surrounding the nanoprotuberances, whereas the lower-frequency component is localized primarily on the nanoprotuberances. This complementary spatial behavior resembles the pattern observed in Figure \ref{fig:MoSe2_1}, possibly indicating a common underlying mechanism such as charge distribution.

\begin{figure}[H]
\centering
\includegraphics[width=0.6\textwidth]{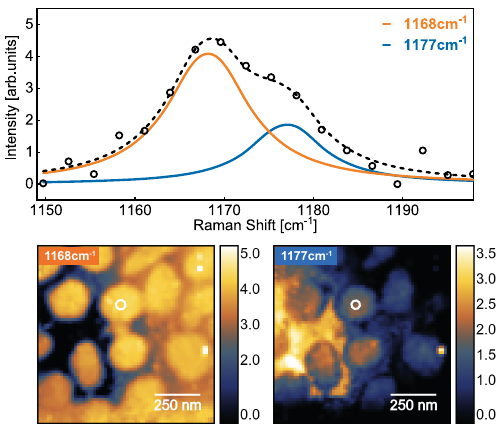}
\caption{Curve fitting for characteristic peaks in nanoprotuberance regions and corresponding TERS intensity maps for HS2 area. The orange Lorentzian corresponds to the peak at 1168cm$^{-1}$ and the blue to 1177cm$^{-1}$. Experimental data are represented by open black circles, while the dashed line corresponds to the final fitted curve. The white circle indicates the pixel in the map from which the spectrum data was extracted.}
\label{fig:MoSe2_3}
\end{figure}

For HS2, the nanoprotuberances become prominent when selecting the peak at 1168 cm$^{-1}$, as shown in Figure \ref{fig:MoSe2_3}. A similar pattern to that observed in HS1 is identified, with multiple Lorentzian components contributing to the signal: the lower-frequency Lorentzian is more pronounced at the locations of the nanoprotuberances, while the higher-frequency component is more intense in the surrounding regions. The peak at 1168cm$^{-1}$ is likely attributed to vibrational modes of organic compounds, such as those involving C-C bonds, which may also indicate the presence of carbon contamination. The map shown in Figure \ref{fig:MoSe2_3} was acquired at the edge of the MoSe$_2$ flake, which accounts for the presence of the observed peak in regions that would otherwise be expected to be flat. It might be due to irregular adhesion of the flake to the underlying hBN or the accumulation of contaminants at the flake boundary at the bottom of the map.

\begin{figure}[H]
\centering
\includegraphics[width=\textwidth]{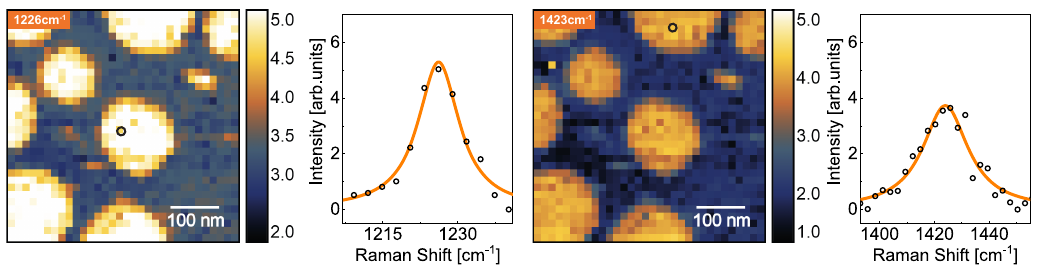}
\caption{Intensity TERS maps of the 1226cm$^{-1}$ and 1423cm$^{-1}$ peaks in nanoprotuberances for HS3 and their corresponding curve fits. Experimental data are represented by open black circles. The black circle indicates the pixel in the map from which the spectrum data was extracted.}
\label{fig:hBN}
\end{figure}

The HS3 hyperspectrum was collected from a region without the presence of MoSe$_2$, where nanoprotuberances are observed over hBN. The maps of the peaks in these regions reveal both similarities and differences compared to the peaks found in HS1 and HS2. In HS3, the peaks most prominently highlighting the nanoprotuberances in intensity maps are at 1226 and 1423cm$^{-1}$, as shown in Figure \ref{fig:hBN}. The first can be attributed to organic compounds (CH$_2$ or C-O-H), and 1423cm$^{-1}$ is associated with CH$_2$ vibration. A single Lorentzian was used to fit the data in these maps.

The presence of a pronounced MoO$_3$ mode near 1000 cm$^{-1}$ reinforces the hypothesis of MoSe$_2$ oxidation, while the detection of vibrational modes attributed to organic compounds in both hBN and MoSe$_2$ regions supports the idea that the nanoprotuberances result from carbon-based contamination. These findings are consistent with the scenario in which thermal cycling under ambient conditions led to the formation of surface contamination protuberances. Interfaces between the contaminants and MoSe$_2$ may exhibit enhanced chemical reactivity, favoring the localized oxidation of MoSe$_2$ and subsequent formation of MoO$_3$. A peak near the known vibrational frequency of molecular oxygen was also detected in HS1 and HS3. Although these features are very weak, they may suggest the presence of trapped gas in the nanoprotuberance regions. However, this possibility remains uncertain and requires further investigation.

\section{Conclusions}

Nano-Raman and AFM analyses were conducted on MoSe$_2$ and hBN regions exhibiting nanoprotuberances. A figure of merit was established to identify the presence of surface contaminants, based on the intensity ratio between characteristic MoSe$_2$ Raman modes (A$_{1g}$/E$_{2g}$). A reduction in this ratio serves as a reliable nano-Raman signature of contamination in monolayer MoSe$_2$ samples. Consequently, TERS measurements can effectively detect the presence of such impurities. In addition, our nano-Raman data revealed spectral features associated with oxidation (MoO$_3$) and contamination (organic compounds) within the nanoprotuberances. These results contribute to a more comprehensive understanding of the nanoscale surface chemistry and degradation processes in two-dimensional materials.

\section{Methods}

The sample herein studied was prepared using a dry-stamping approach. Initially, an hBN flake was lifted with a polymer stamp. The MoSe$_2$ grains, grown by the chemical vapor deposition (CVD) method, were then deposited on the hBN surface (MoSe$_2$/hBN). Finally, the assembled structure was transferred to a glass slide using another polymer stamp, ensuring proper layer alignment while avoiding direct contact with the SiO$_2$ substrate (coverslip) \cite{masubuchi2022dry,naito2023high}. This method may result in wrinkles and protuberances between the MoSe$_2$ and hBN layers, or between hBN and SiO$_2$ substrate, once the application of high pressure to the stamp induces stresses within the layer, which may lead to nanoprotuberance formation. Imperfect material cleaning and further exposure to air can also generate contaminations. 

For this nano-Raman spectroscopy study, we employed tip-enhanced Raman spectroscopy (TERS) based on an atomic force microscope (AFM) to analyze the sample. TERS measurements were performed using the Porto Laboratory prototype system, which operates in bottom illumination mode with a non-contact AFM setup utilizing a tuning fork \cite{Rabelo_2019}. The system is equipped with a He-Ne radially polarized as the excitation source, and both an avalanche photodiode (APD) detector and an Andor Shamrock 303i spectrometer with a 600 l/mm grating. The TERS probes used in this experiment were PTTP (Plasmon Tunable Tip Pyramids) probes, specifically chosen for their ability to enhance Raman scattering through localized surface plasmon resonance \cite{vasconcelos2015tuning, Vasconcelos_2018, Miranda_2020}. The nano-Raman hyperspectra were recorded with a step size of 15.6nm. The HS1 region was analyzed using a tip that provided a 25-fold enhancement for the A$_{1g}$ mode (tip in/ tip out contrast), while the remaining two (HS2 and HS3) were analyzed with a tip that offered a 10-fold spectral enhancement for the same Raman peak.

The acquired data was processed using PortoFlow Analysis software, where Principal Component Analysis (PCA) was applied to improve the data quality. PCA transforms the data into a lower-dimensional space composed only by five principal components as reconstructed data, highlighting the most relevant information and enabling the identification of key features in the Raman spectra. This approach improved the clarity of the spectral data, particularly in visualizing the intensity and spatial distribution of weak Raman peaks across the sample. The spectral regions of interest were selected, and background was removed to allow proper curve fitting. This procedure produced intensity maps based on the properties of the Lorentzian peaks used to fit the spectral features.

For completeness, AFM measurements were performed using a Park Systems XE-70 AFM operating in tapping mode. The AFM analysis was crucial for providing AFM-standard topographical information of the sample, allowing us to correlate structural features with the Raman data obtained from the TERS measurements. 

\begin{acknowledgement}

The authors thank financial support by FAPEMIG (APQ - 04852-23, APQ - 01860-22, RED - 00081-23, APQ-01402-23, RED-00079-23), the Japan Science and Technology Agency (JST), the JST FOREST Program (JPMJFR213X), the CREST (JPMJCR24A5), Kakenhi Grants-in-Aid (JP21H05232, JP21H05233, JP21H05234,  JP22H00283, JP22H04957, and JP23H02052) from the Japan Society for the Promotion of Science (JSPS), World Premier International Research Center Initiative (WPI), MEXT, Japan, and software resources and technical assistance provided by FabNS. R.S. acknowledges a JSPS KAKENHI Grant (No. JP22H00283), Japan and the Yushan Fellow Program by the Ministry of Education (MOE), Taiwan.

\end{acknowledgement}

\providecommand{\latin}[1]{#1}
\makeatletter
\providecommand{\doi}
  {\begingroup\let\do\@makeother\dospecials
  \catcode`\{=1 \catcode`\}=2 \doi@aux}
\providecommand{\doi@aux}[1]{\endgroup\texttt{#1}}
\makeatother
\providecommand*\mcitethebibliography{\thebibliography}
\csname @ifundefined\endcsname{endmcitethebibliography}  {\let\endmcitethebibliography\endthebibliography}{}

% For two cols:
%\end{multicols}{}
%%%%%%%%%%%%%%%%%%%%%%%%%%%%%%%%%%%%%%%%%%%%%%%

\end{document}